# Ionization waves (striations) in a low-current plasma column revisited with kinetic and fluid models

**J.P. Boeuf [a]**

LAPLACE, Université de Toulouse, CNRS, INPT, UPS, 118 Route de Narbonne, 31062 Toulouse, France

**ABSTRACT**

A one-dimensional Particle-In-Cell Monte Carlo Collisions (PIC-MCC) method has been used to model the development and propagation of ionization waves in neon and argon positive columns. Low current conditions are considered, i.e. conditions where stepwise ionization or Coulomb collisions are negligible (linear ionization rate). This self-consistent model describes the development of self-excited moving striations, reproduces many of the well-known experimental characteristics (wavelength, spatial resonances, potential drop over one striation, electron "bunching" effect) of the ionization waves called p, r and s waves in the literature and sheds light on their physical properties and on the mechanisms responsible for their existence. These are the first fully kinetic self-consistent simulations over a large range of conditions reproducing the development of p, r and s ionization waves. Although the spatial resonances and the detailed properties of the striations in the non-linear regime are of kinetic nature, the conditions of existence of the instability can be obtained and understood from a linear stability analysis of a three-moment set of quasi-neutral fluid equations where the electron transport coefficients are expressed as a function of electron temperature and are obtained from solutions of a 0D Boltzman equation. An essential aspect of the instability leading to the development of these striations is the non-Mawellian nature of the electron energy distribution function in the uniform electric field prior to the instability onset, resulting in an electron diffusion coefficient in space much larger than the energy diffusion coefficient.

## I. INTRODUCTION

Plasma columns (e.g. positive column of a glow discharge) often exhibit striations of the emitted light, i.e. the alternance of dark and luminous regions that are immobile or moving. The various states of plasma columns in rare gases (diffuse, striated, constricted) have been known since the middle of the 19th century. A considerable number of papers have been devoted to experimental and theoretical studies of these striations (see, e.g., the reviews of Refs. [1-6]) but, due to the large variety of regimes and to the complexity of the physics involved, our understanding of these instabilities is limited. Striations form in molecular gases, in electronegative gases as well as in inert gases and the nature of the gas certainly plays an essential role in the dispersion relations and in the development of these instabilities. The dispersion relations of a plasma column[7] can be quite complex if one takes into account all possible hydrodynamic instabilities, as has been done by Haas[8, 9] in a paper described as "formidable" by Allis in his review article[9], where he wrote and linearized the three conservation equations for a 5-component plasma including neutral molecules, metastables, electrons, positive ions, and negative ions.

In this paper we focus on a particular type of striations that can develop in inert gases in the absence of Coulomb collisions and stepwise ionization and where charged particle losses are due to recombination on the walls.

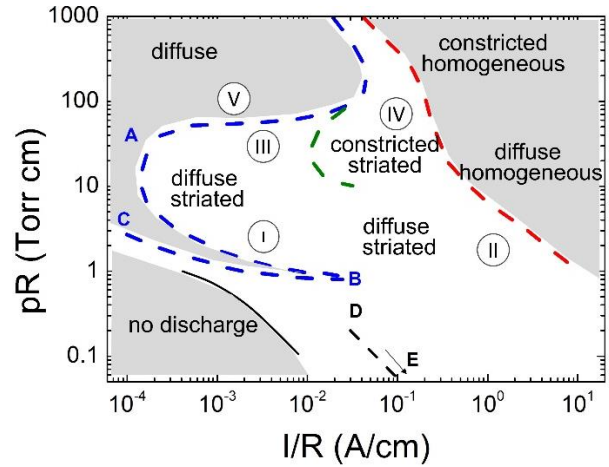

Figure 1: Schematic of the different regimes of a positive column plasma in neon (after Refs.[6], [4]). See text. The regime studied in this paper corresponds to the low pressure low current striated regime (region I or region slightly below the BC line).

In inert gases, the different regimes of striations can be mapped in the $(pR, I/R)$ diagram of Figure 1, where $p$ is the gas pressure, $R$ the discharge radius, and $I$ the discharge current. Striations exist in the non-greyed region of this map. For large $I/R$, on the right of the red dashed line called the Pupp boundary[2, 10], the plasma is homogeneous (striation free), constricted for large values of $pR$, and diffuse for lower

---

values. Striations appear on the left of the Pupp boundary in a diffuse (region II), or constricted form (region IV). The plasma column is homogenous on the left of the dashed blue line of Figure 1 (region V) including in the "tongue" region between lines AB and BC. Region III is characterized by the development of irregular striations while three well-defined types of striations (termed as *p*, *r*, and *s* striations) are present in region I. The DE region corresponds to capillary discharges[4].

The plasma instabilities can take the form of "standing striations" or "moving striations". In region I in inert gases the phase velocity of ionization waves is generally directed from anode to cathode while the group velocity is directed towards the anode (backward waves).

The non-linearity of the ionization rate, i.e. its non-linear dependence on the electron density due to Coulomb collisions (leading to a Maxwellization of the electron energy distribution function) or to stepwise ionization is known to play an important role in a large part of the domain of existence of striations represented in Figure 1 (region of sufficiently large values of $I/R$ and $pR$). It is generally admitted that hydrodynamic models, i.e. models based on fluid equations are able to describe the formation of striations in this domain. Plasma stratification in a radiofrequency discharge in argon, for example, can be simulated successfully under conditions close to the experiments, with a two-dimensional fluid model assuming a non-linear ionization rate[11].

In the conditions considered in the present paper, i.e. at low pressure and low current (region I of Figure 1) in inert gases, Coulomb collisions and stepwise ionization are less important and the non-linearity of the ionization rate is no longer the dominant cause of plasma stratification. In these conditions, the electron distribution function is not Maxwellian, and the nature of the striations has been attributed to kinetic effects[4-6, 12]. An important feature of the striations in this regime, which was first revealed in the experiments of Novak[13], is the fact that the potential drop over the length $\lambda$ of a striation, $\delta\phi_\lambda$, is practically independent of the discharge conditions and mainly determined by the type of gas. The value of this constant characterizes the type of wave (*p*, *r*, or *s* wave) and is related to the direct excitation thresholds, $U_{exc}$, of the specific gas (on the order of 20 eV for Ne and 12 eV for Ar), with the following approximate law: $\delta\phi_\lambda{}^s \approx U_{exc}$, $\delta\phi_\lambda{}^p \approx U_{exc}/2$ and $\delta\phi_\lambda{}^r \approx 2U_{exc}/3$ for *s*, *p*, and *r* waves respectively. In the experiments of Figure 2[14] no striations were observed for currents below 5 mA, and *p*, *r*, and *s* striations appeared sequentially with hysteresis loops when increasing the current above 5 mA. When the $pR$ product decreases, the charged particle losses to the walls increase and the electric field in the positive column must increase. Figure 3 shows the variations of the wavelengths of *p*, *r*, and *s* waves in neon as a function of the axial electric field $E_0$, measured by Novak[13] for different tube radii and neon gas pressure.

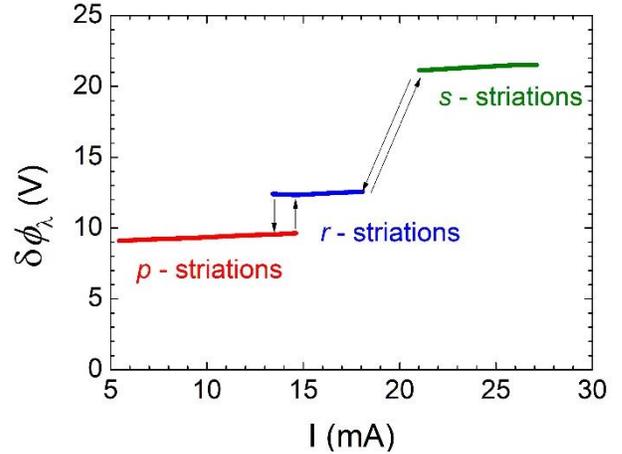

Figure 2 : Measured potential drop $\delta\phi_\lambda$ over the striation length as a function of discharge current in neon at 1.55 torr, tube radius R=1 cm. After Golubovskii et al.[14].

$E_0$ is the absolute value of the average electric field in the column. The total longitudinal electric field at a given position $x$, $E(x)$, is the sum of the average field and of the space charge electric field $E_{sp}$ (related to the density gradients in the striation). Since $E_0$ is the average electric field, the potential drop over one striation, $\delta\phi_\lambda$, can be written as:

$$\delta\phi_\lambda = \left| -\int_{striation} E(x)\, dx \right| = E_0\lambda \qquad (1)$$

We see in Figure 3 that the wavelength decreases with increasing electric field (decreasing $pR$) and increases with the nature of the striations from *p* to *r* and *s*.

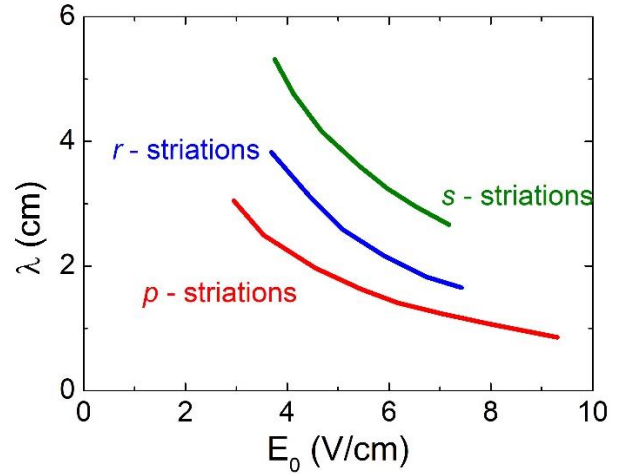

Figure 3: Striation lengths $\lambda$ for the *p*, *r*, and *s* waves, as a function of average longitudinal electric field $E_0$, measured in a neon positive column plasma in discharge tubes of diameters 1.12, 1.73, and 2.14 cm, for pressures between 1 and 5.5 torr, and for currents from 0.2 to 8 mA. After Novak[13]. Note that these curves are independent of pressure if one assumes that the potential drop between striations is constant for each type of wave (see Figure 2 and text), i.e. $\delta\phi_\lambda = E_0\lambda = const$.

Note that the reasons for the dependence of the striation type on the current in Figure 2 is not completely clear. The authors of Ref.[6] indicate that the electric field weakly depends on the current (this could be due to the effect of non-negligible stepwise ionization) and that hysteresis effects are present. In our models and simulations the only control parameter is the average electric field $E_0$ since we neglect all non-linear effects with plasma density and gas pressure.

To study theoretically the possible role of non-hydrodynamic or kinetic effects in the development of $p$, $r$, and $s$ striations, Ruszicka and Rohlena[15] linearized a two-term Legendre expansion form of the Boltzmann equation assuming a given, spatially periodic perturbation of the electric field. They found that the amplitude of the response of the electron energy distribution function (EEDF) and electron density to this perturbation was maximum when the potential drop over one striation was of the form $\delta\phi_\lambda \approx U_{exc}/q$, where $U_{exc}$ is the excitation threshold (assuming a single electron excitation level and neglecting elastic energy losses), with $q$ being an integer, $q = 1,2, ...$This was consistent with the experimental results of Novak for the $s$ ($q = 1$) and $p$ ($q = 2$) waves.

The consequence of this resonant properties for potential drops of the form $\delta\phi_\lambda \approx U_{exc}/q$ across the striation is the existence of bumps in the EEDF along the striation length (electron "bunching" effect). Low energy electrons in the low field region of a striation are accelerated to a potential close to $\delta\phi_\lambda$ between two striations, leading to maxima of the EEDF at $\delta\phi_\lambda$ (and possibly at multiples of $\delta\phi_\lambda$).

A more general analytical theory of the relaxation of the EEDF in spatially periodic electric field was developed by Tsendin[16]. This theory confirmed the presence of resonance peaks for the electron distribution function for $\delta\phi_\lambda \approx U_{exc}/q$ [4, 5, 16, 17]. Shveigert[18] performed computational analysis of the kinetic equation of Tsendin in slightly modulated electric field and confirm the existence of these spatial resonance phenomena. Other numerical solutions of the Boltzmann equations in spatially periodic electric fields by Golubovskii et al.[14, 19] confirmed the resonances for $q = 1$ and $q = 2$ corresponding to the $s$ and $p$ striations and evidenced the presence of a resonance for $\delta\phi_\lambda = 2/3\ U_{exc}$ corresponding to the experimentally observed $r$ striations.

Prior to the theoretical approaches mentioned above, experiments by Rayment[20] had shown the presence, in the electron energy distribution function (EEDF), of striated low current columns, of bumps, moving along the striation length. These structures in the energy distribution function were theoretically interpreted by Tsendin as a result of «bunching» of the EEDF[16, 17], the resonances described above being a « manifestation of the EEDF bunching effect »[5]. This electron bunching effect was later confirmed in other experiments and in several papers on solutions on the Boltzmann equation in perturbed, periodic electric fields (see, e.g. Ref.[21] and the review papers Refs.[4-6]).

Recently, a self-consistent hybrid simulation of dc and rf positive column plasmas in argon was presented[22] under conditions where the effect of Coulomb collisions and stepwise ionization were negligible. The electric field was not imposed but rather obtained from Poisson's equation, electron transport was described kinetically with a Monte Carlo simulation, and continuity and drift-diffusion equations were used for positive ions. The simulation was able to describe the formation of striations with properties comparable to experiments. The potential drop along one striation was on the order of 10 eV in argon (probably $r$ wave). Finally a self-consistent hybrid model of striations in low-current discharges based on a Fokker-Planck kinetic equation for electrons has been recently submitted for publication by Kolobov et al.[23].

In this paper we present results from a fully-kinetic self-consistent model (PIC-MCC simulation) of low-pressure, low-current positive column plasmas in argon and neon in a large range of conditions (electric field, or positive column radius). The model reproduces the formation of striations with properties similar to the experimentally observed ionization waves. We show that the conditions and the reasons for the existence of this instability can be understood from the linearization of a three-moment quasineutral fluid model similar to the one recently presented by Désangles et al.[24]. The wavelength and growth rate predicted by the linearization of the fluid model follow the same trends as those deduced from the PIC-MCC simulations, but the fluid model cannot predict the characteristic spatial resonances of these ionization waves.

The PIC-MCC model is described in section II. Results from this kinetic model are presented in section III. The linear stability analysis of a quasineutral fluid model of the plasma column is detailed in section IV.A and section IV.B show comparisons between the wavelength, growth rate and phase velocity of the wave predicted by this model and those deduced from the PIC-MCC simulations. In the last section, V, we make a few remarks on periodic vs bounded plasma models.

## II. PIC-MCC MODEL

PIC-MCC methods are based on the simulation of the trajectories of a large number of electrons and ions in phase space under the effect of the self-consistent electric field and of collisions with neutral atoms. The electric field is calculated at each time step knowing the charged particle densities on a spatial grid, which is deduced by interpolation of the particle positions. There are constraints on the time step ($\delta t < 0.2/\omega_{pe}$, where $\omega_{pe}$ is the electron plasma frequency) and on the grid spacing ($\delta x < \lambda_{De}$, where $\lambda_{De}$ is the electron Debye length)[25]. Electron-neutral and ion-neutral collisions are simulated with a Monte Carlo method where the probability of collisions, the nature of collisions and the scattering after collisions are deduced from the scattering cross-sections.

In this paper we used the 1D (one-dimensional) module of a user-friendly 2D Open MP parallelized PIC-MCC simulation code, J-PIC. The 2D version J-PIC has been used in several recent papers on the simulation of magnetized plasmas (see, e.g., Refs.[26-32] ). J-PIC is based on standard features of the electrostatic PIC-MCC method as described by Birdsall[25]. The 1D module of J-PIC has been validated by comparisons with several published papers including the benchmark paper of Turner et al.[33] and the results of eduPIC presented in the article by Donko et al.[34].

Results are presented below for argon and neon. The electron-neutral cross-sections used in the simulations are taken from the siglo database[35] of LXCat[36]. The argon electron-neutral cross-sections correspond to a compilation of A.V. Phelps where the electron excitation levels are grouped in one single level, while the neon cross-sections are the same as those described in Meunier et al.[37] and include seven excitation levels. The isotropic and backward scattering ion-neutral cross-sections for argon are taken from Phelps[38], those for neon are taken from Piscitelli et al.[39]. The sharing of energy between the scattered and ejected electrons after ionization is the same as in Refs.[34, 40].

To simulate a positive column in a diffusion dominated regime, i.e. when charged particle losses are due to radial ambipolar diffusion and recombination at the walls we used the simple 1D model described below. To avoid the cumbersome description of the cathode region of a glow discharge we can consider a given length $L$ of the plasma column and assume periodic boundary conditions (electrons or ions reaching one boundary are reinjected with the same velocity at the other boundary; the electric potential is deduced from Poisson's equation by solving two tridiagonal systems based on the Thomas algorithm[41]). An average longitudinal electric field $E_0$ is imposed in the column so that the total electric field $E(x)$ at a given location $x$ is the sum of the applied field and of the space charge electric field $E_{sp}$ deduced from Poisson's equation: $E = -E_0 + E_{sp}$ (the applied field in the simulations is negative and as said above $E_0$ the absolute value of the applied field). In a diffusion dominated positive column ionization is exactly balanced by charged particle losses to the walls. Therefore, each time an ionization takes place, a new electron-ion pair is generated, and another electron and ion are chosen randomly and separately over the simulation length, and are removed from the simulation. Doing so, the total number of particles remain constant. The simulation is started with uniform and equal electron and ion densities and the space averaged electron and ion densities remain constant during the simulation. In this model the frequency of charged particle losses to the walls (equal to the ionization frequency) is related to the applied field $E_0$ and, assuming ambipolar losses to the walls, the positive column radius $R$ is imposed by $E_0$ through the usual relation:

$$\nu_i(T_e) = (2.4/R)^2 D_a(T_e) \qquad (2)$$

where $T_e$, $\nu_i(T_e)$, and $D_a(T_e) = D_e \mu_i/\mu_e$ are respectively the electron "temperature" (2/3 of the electron mean energy), ionization frequency, and ambipolar diffusion coefficient and $D_e$, $\mu_e$, $\mu_i$ are the electron free diffusion coefficient, the electron mobility and the ion mobility. In the homogeneous regime, the electron transport coefficients and collisions frequencies, and their dependence on electron temperature can be obtained from solutions of the homogeneous and steady state Boltzmann equation for different values of the electric field $E_0$.

The results of sections III, IV, and V have been obtained at a given pressure (1 torr) and for different values of $E_0$. Since stepwise ionization and Coulomb collisions are neglected, the usual discharge scaling laws are valid. Some of the results are therefore presented in reduced units such as $E_0/p$, $\nu_i/p$, $Rp$,

$\lambda p$ (wavelength times pressure) , and $\gamma/p$ (growth rate over pressure). Finally, the model being linear in plasma density, the results are independent of the value of the initial (or space averaged) plasma density in the simulation. We therefore chose relatively low values of the plasma density (to avoid strong limitations on the time step) but large enough so that the electron Debye length was always much smaller than the wavelength of the instabilities. The plasma density was $10^{15}$ m$^{-3}$ in all simulations (at a pressure of 1 Torr, gas temperature 300 K, and for column lengths of 4 to 15 cm). The number of charged particles per cell was more than 500 in all simulations.

## III. PIC-MCC SIMULATION RESULTS

### A. General properties of the striated regime

Figure 4 shows a typical example of the striated regime obtained with the self-consistent PIC-MCC model for a periodic plasma column of length $L = 6$ cm in neon at a pressure $p = 1$ torr. The applied field is negative (the cathode would be on the left side of the figure) and its absolute value $E_0$ is 6 V/cm. The large amplitude oscillations of the different quantities displayed in Figure 4 are moving without deformation towards the cathode side. This quasi-stationary regime is reached in about 40 μs in the conditions of Figure 4 but the time necessary to reach a steady state increases with decreasing applied electric field (the growth rate of the instability and its variations with $E_0$ are discussed in section IV).

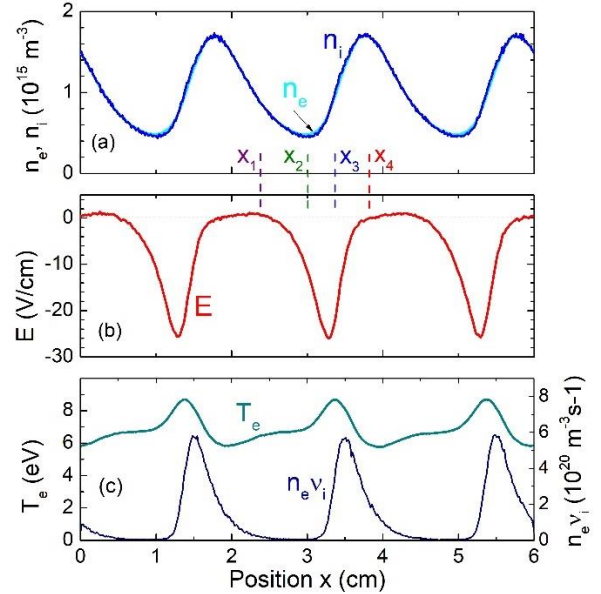

Figure 4: (a) Electron and ion densities, (b) electric field and, and (c), electron temperature and ionization rate profiles in a plasma column of length $L = 6$ cm in neon at a pressure $p = 1$ torr for an applied field $E_0 = 6$ V/cm (i.e. averaged field -6V/cm) obtained with the PIC-MCC simulation (the model is linear in electron density so the usual scaling laws with pressure can be used). The ionization wave is moving to the left . Uniform initial plasma density $n_e = n_i = 10^{15} m^{-3}$, number of grid points 400, 1000 electrons and ions per cell. The calculated current density is about 10 A/m$^2$.

The total electric field in Figure 4b is the sum of a resistive term and an ambipolar component related to the plasma density gradients. The ambipolar field is such that the total electric is very small and even slightly positive in regions of maximum plasma density, i.e. electrons are trapped in these regions and are accelerated by the large electric field between these maxima.

The plasma density maxima are located in the low field region and the maxima of the electric field are shifted towards the minima of the electron density to ensure current continuity.

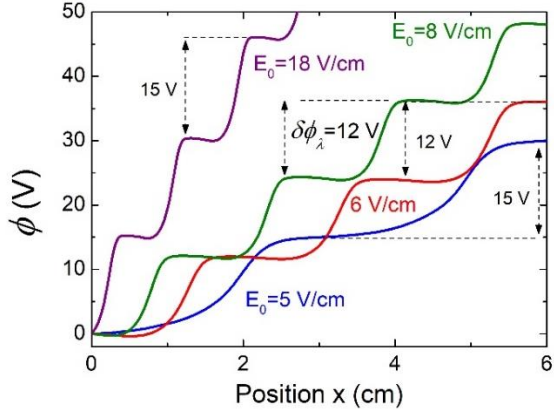

Figure 5: Axial profile of the electric potential for different values of the longitudinal electric field $E_0$ (5, 6, 8, 18 V/cm) in a plasma column in neon, $L = 6$ cm and $p = 1$ torr, average plasma density $10^{15} \text{m}^{-3}$.

The electron temperature (Figure 4c) is modulated by about ±1 eV around its average value, with maxima slightly shifted with respect to the electric field maxima. The ionization rate also shown in Figure 4c is considerably more modulated than the electron temperature and its maxima are shifted in the direction of electron motion with respect to the electric field and electron temperature. This is expected since the energy relaxation length of the low energy electrons (which controls the electron temperature) is much larger than that of the electrons in the tail of the energy distribution function (which controls ionization).

Figure 5 displays the axial profiles of the electric potential for different values of the electric field $E_0$ obtained for the PIC-MCC simulation of a 6 cm plasma column in neon at 1 torr. The potential drop between striations is between 12 and 15 V in the examples of Figure 5.

### B. Wavelength and potential drop over a striation

We performed simulations of the striated regime in a large range of variations of the average electric field in neon and argon positive columns. We show in this sub-section the calculated variations of the wavelength and potential drop over a striation as a function of $E_0$ (see also section IV.B). $E_0$ is imposed in the simulations, but is actually imposed in the experiments by the charged particle balance equation, eq. (2) and is therefore related to the tube radius. We discuss below the relation between $E_0/p$ and $pR$ for neon and argon.

The calculated potential drop and wavelength of the striations are plotted as a function of $E_0$ in Figure 6 for neon and argon.

Although the potential drop is not perfectly constant as seems to be the case in the experiments reported, for example, by Golubovskii et al.[14], we clearly see two distinct groups of values of the potential drop for each gas. In neon, the two groups of values of the potential drop between striations on the order of 13 V (between 10 V and 16 V) and around 23 V (between 22 and 26 V). In argon the two groups are around 11 V and 16 V. This seems to correspond to ionization waves of types $r$ and $s$ respectively (see Figure 2 for neon, and Refs.[2, 13, 14]). For some of the lowest values of $E_0$ in neon, the potential drop is about 10 V, which is close to the potential of drop of $p$ waves. In both cases (neon and argon) the $r$ type was dominant for simulations with lower values of the electric field (larger tube radii) while the $s$ type took place at larger values (above 20 V/cm). The striation length $\lambda$ of Figure 6b obtained from the PIC-MCC simulations is of course consistent with the potential drop $\delta\phi_\lambda$ of Figure 6a and satisfies eq. (1): $\delta\phi_\lambda = E_0\lambda$ .

Starting the simulation with a uniform initial plasma density the formation time of the instability was very large, on the order of several hundreds of microseconds, at 1 torr for the smallest values of the electric field in Figure 6 and decreased by about two orders of magnitude for the largest values of $E_0$.

This can be understood by the large variation of the ionization frequency (and wall loss frequency) in this range (see Figure 7). The question of the growth rate of the instability is discussed in section IV. No instabilities were found below $E_0 = 4.3$ V/cm in neon and $E_0 = 11$ V/cm in argon. This could be due to physical boundaries of the instability domain, or to a very large characteristic growth time of the instability (difficult to describe in a reasonable computation time). This question is also discussed in section IV.

Each value of $E_0$ in the PIC MCC simulations corresponds to a value of the positive column radius $R$ through the usual charged particle balance relation eq. (2), leading to:

$$pR = 2.4\sqrt{pD_a/(v_i/p)}$$

where the reduced ambipolar diffusion coefficient $pD_a$ and the ionization frequency $v_i/p$ can be calculated as a function of $E_0/p$ using a Boltzmann solver.

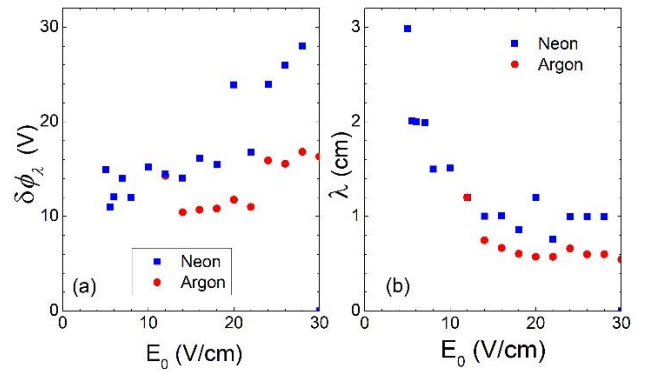

Figure 6: (a) Potential drop $\delta\phi_\lambda$ between striations, and (b), striation length as a function of the plasma column electric field $E_0$ in neon and argon from PIC-MCC simulations on neon and argon for $p = 1$ torr , $L = 6$ cm. The relation between $E_0$ and the column radius is shown in Figure 7a.

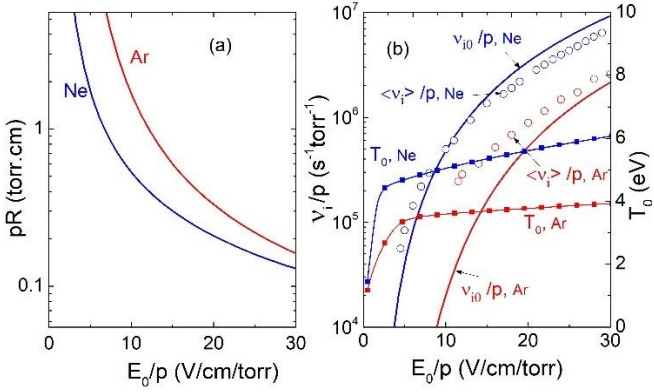

Figure 7: (a) $pR$ product in neon and argon ($R$ is the radius of the positive column) as a function of $E_0/p$ (at 300 K) deduced from the charged particle balance equation (2), with ionization frequency and ambipolar diffusion coefficient calculated with the 0D Boltzman equation solver BOLSIG+[42], (b) Electron temperature $T_0$ and reduced ionization frequency $\nu_{i0}/p$ as a function of $E_0/p$ in neon and argon calculated with BOLSIG+ for a uniform field (the PIC-MCC simulations under a uniform field give the same results). The open circle symbols correspond to the ionization rate calculated with the PIC-MCC simulation in the non-linear, staturated regime, and averaged over the simulation length. The cross-sections for electron collisions with neon and argon used in the Boltzman solver and in the PIC-MCC simulations, are described in section II . The ion mobility is taken from Ellis et al.[43].

The variations of $pR$ with $E_0/p$ obtained for neon and argon with the Boltzmann solver BOLSIG+[42] (i.e. under a uniform electric field) are shown in Figure 7a and the corresponding ionization frequency and electron temperature are displayed in Figure 7b.

The large values of $E_0/p$ correspond to small tube radii (e.g. capillary tubes at pressures around 1 torr) or to low pressure (e.g. radii in the cm range and pressures around 100 mtorr), while the lowest values of $E_0/p$ in the simulations above corresponds to more standard situations (e.g. cm radii, pressure in the torr range). Note that in the striated regime, the field is not uniform and thus the variations of the average $\nu_i/p$ and of $pR$ with $E_0/p$ are different from those obtained in a uniform field. Figure 7b shows, for comparisons the space averaged ionization rate in the non-linear regime obtained with the PIC-MCC simulation. The difference between uniform field and non-linear regime are larger for argon, especially at low electric fields.

### C. Electron energy probability function

As mentioned in the introduction, previous experimental and numerical works have shown that the electron energy distribution function can present bumps that move along the striation length. The PIC-MCC model can provide detailed information about the shape and variations of the self-consistent electron energy distribution.

Figure 8 shows the normalized electron energy probability functions (EEPF) in log scale in the conditions of Figure 4, at four locations (indicated by $x_1, x_2, x_3, x_4$ in Figure 4). The applied field $E_0$ in these conditions was 6 V/cm and the calculated potential drop was 12 V (see Figure 6a), which seems to corresponds to an $r$ wave according to the theory (2/3 of the excitation thresholds, which are around 19-20 eV in neon) and to the experimental results of Figure 2.

In the representation of Figure 8, a Maxwellian distribution would be a straight line. The normalization of the EEPF, $F(x,\varepsilon)$, is such that $\int \varepsilon^{1/2} F(x,\varepsilon)\,d\varepsilon = 1$. The EEPF calculated for a uniform field $E_0 = 6$ V/cm is also shown for comparison. The four positions $x_1, x_2, x_3, x_4$, correspond respectively to the end of the low field region on the side of the negative density gradient (i.e. anode side), to the region of minimum plasma density and acceleration by the electric field, to the end of the acceleration region, and to the maximum plasma density in the low field region.

In the low field regions around $x_1$ and $x_4$, we see a large group of low energy electrons. These electrons are trapped by the small relative maximum of the plasma potential in the low field region. The trapping effect is enhanced by a lower electron-neutral momentum cross-section (neon) at low energy or by the presence of a Ramsauer minimum (argon) and has been observed, for example, in recent measurements by Godyak et al.[44] of the EEPF in low pressure striated positive columns in argon (we observe the same effect in the PIC-MCC simulations in argon). The EEPF at location $x = x_1$ presents a bump around 12 eV and a depleted tail. From $x_1$ to $x_2$ the field has increased from 0 to about 10 V/cm, corresponding to a potential increase of about 2 V. We see that the bump seen at the position $x_1$ has moved to 14 eV while another bump appears around an energy of about 2 eV. This energy shift is clearly associated with the potential increase. The low energy peak at $x_1$ has disappeared and is now around 2 eV, i.e. the trapped electrons of location $x_1$ have been released and gained an energy of 2 eV.

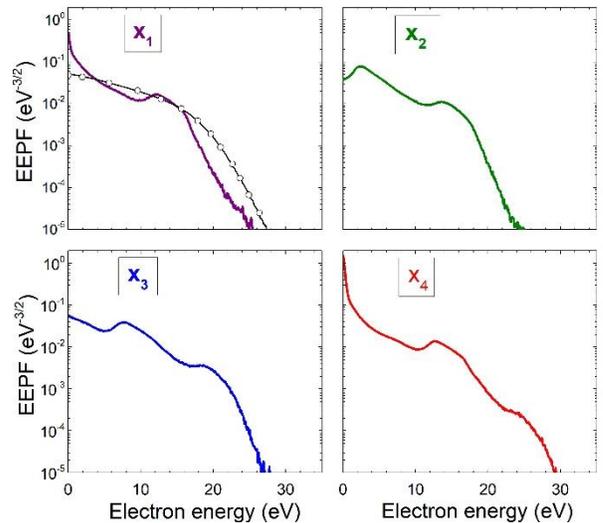

Figure 8: Normalized electron energy probability function, EEPF, $F(x,\varepsilon)$ in the conditions of Figure 4 ($E_0 = 6$ V/cm , neon, $p = 1$ torr , $L = 6$ cm) at four different positions (noted $x_1, x_2, x_3, x_4$ in Figure 4). The black line with open circle on the top left plot, shows, for comparison, the EEPF calculated in a uniform field $E_0 = 6$ V/cm (black line : PIC-MCC, open circles : Boltzmann solver BOLSIG+).

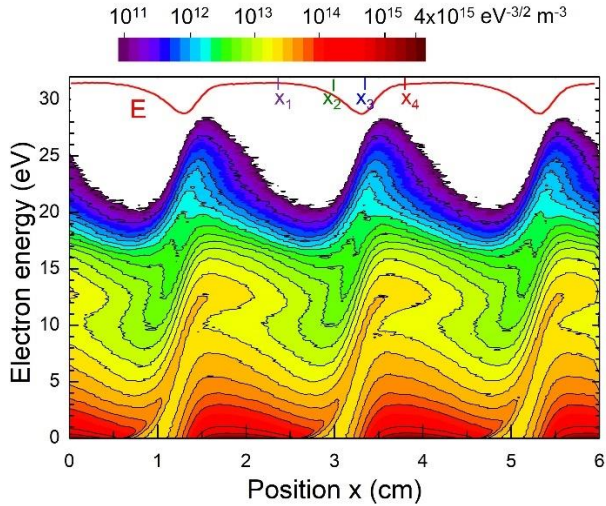

Figure 9: Contours of constant EEPF normalized to the plasma density (i.e. $f(x, \varepsilon) = n_e(x)F(x, \varepsilon)$ ) in the conditions of Figure 4 and Figure 8 ($E_0 = 6$ V/cm , neon, $p = 1$ torr, $L = 6$ cm). The electric field profile is shown on top of the contour plot. The positions $x_1, x_2, x_3, x_4$ of the normalized $F(x, \varepsilon)$ EEDF of Figure 8 are indicated by vertical bars on the top axis.

Elastic energy losses are negligible in these conditions and the low energy region is not repopulated at $x_2$ because of the depleted tail at $x_1$ (i.e. inelastic collisions and ionization are not important between positions $x_1$ and $x_2$ as can be seen on Figure 4c).

From $x = x_1$ to $x = x_3$ the two bumps of the EEPF continue to move to higher energy 8 and 18 eV respectively) because the potential continues to increase. The energy bump at 18 eV is not as visible as the lower energy one and the EEPF actually forms a plateau around 18 eV before decreasing for higher energies. Inelastic collisions start to re-populate the low energy part of the distribution.

At the position $x = x_4$, which is close to the location of the maximum plasma density in the low field region the distribution is close to that at $x_1$ except that the population of the tail of the distribution is more important around $x_4$. The lower energy bump, now at 12 eV, is clearly apparent while the higher energy one, now around 22 eV is much less marked. The tail of the EEPF becomes depleted between $x_4$ and the next location equivalent to $x_1$ ($x_1$ shifted by one wavelength, i.e. $x_1 + \lambda$) because of energy losses due to inelastic collisions and no energy gain in the quasi-zero electric field. The higher energy bump therefore disappears between location $x_4$ and $x_1 + \lambda$, but the 12 eV energy bump is still present.

Figure 9 presents a complete view of the EEPF in the plane $(x, \varepsilon)$, in the form of contour plot. The EEPF is normalized to the plasma density, $f(x, \varepsilon) = n_e(x)F(x, \varepsilon)$. We recognize the features described for Figure 8. The modulation of the tail of the EEPF associated with the oscillations of the electric field and responsible for the modulation of the ionization rate of Figure 4c is clearly visible.

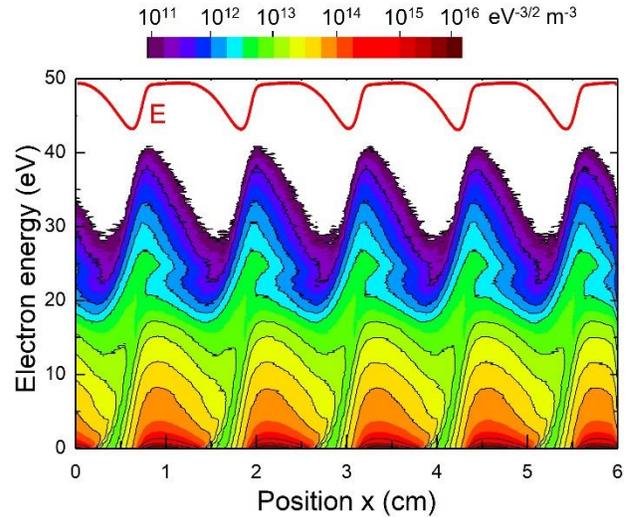

Figure 10: Contours of constant EEPF normalized to the plasma density in neon at $p = 1$ torr and for $E_0 = 20$ V/cm and $L = 6$ cm. The electric field profile is shown on top of the contour plot (minimum at -75 V/cm to be compared with -28 V/cm in the case of Figure 4, 8 and 9).

The formation of the bump in the EEPF in the large field region between location $x_2$ and $x_4$, due to the un-trapping and acceleration of the low energy electrons appears as the formation of an electron beam in this representation. The second bump or plateau around 22 eV is scarcely visible on this plot.

The shape of the EEPF in the striated regime depends on the average electric and on the potential drop along a striation.

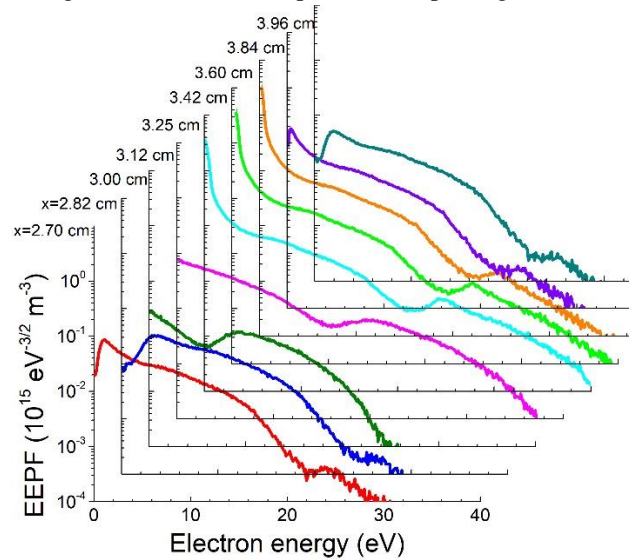

Figure 11: EEPF normalized to the electron density at different locations around one striation, between $x = 2.70$ cm and $x = 3.96$ cm. The conditions are the same as in Figure 10 (neon , $p = 1$ torr, $E_0 = 20$ V/cm and $L = 6$ cm). The striation length is 1.2 cm, the large field region between striations is between $x = 2.70$ cm and $x = 3$ cm, and the high density low field region is between $x = 3$ and $x = 3.9$ cm.

For the larger average field of Figure 10 and Figure 11 ($E_0 = 20$ V/cm) the potential drop along a striation is 22 eV (close to the ionization threshold), which corresponds, according to the theory and to the experimental results (Figure 2) to an *s* wave. As in the previous case of Figure 9, we see a large population of trapped low energy electrons in the regions of maximum plasma density. These electrons are un-trapped and accelerated by the large electric field in the regions between maxima of the plasma density and reach an energy around 22 eV, consistent with the potential drop. In the acceleration region the energy gain due to the electric field is clearly larger than the energy losses due to collisions as can be seen on the first two curves of Figure 11 ($x = 2.70$ cm and $x = 2.82$ cm)

## D. Velocity and frequency of the ionization wave

In all the simulations, the striations were moving in the direction of the electric field, i.e. from anode to cathode. The motion of the striations can be understood by considering the phase-shift between plasma density and electric field and is illustrated in Figure 12. This phase-shift is due to current continuity i.e. the resistive part of the electric field tends to increase when the electron density decreases. The ionization rate is slightly shifted with respect to the electric field (in the direction of electron motion) and is maximum between the maximum electric field and the maximum plasma density. Therefore, new electrons and ions are generated ahead of the maximum plasma density in the direction opposite to electron motion and the structure moves in this direction. This qualitative description clearly shows that the moving striations are "ionization waves".

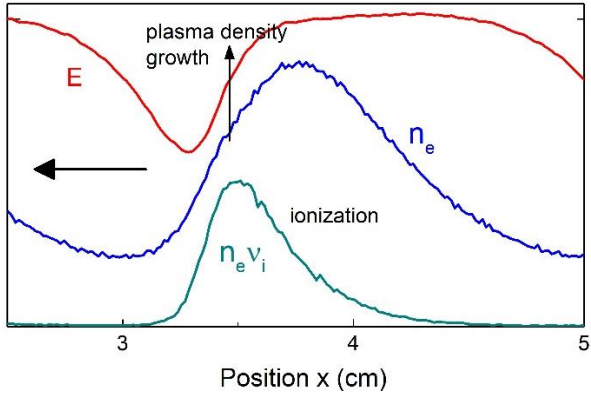

Figure 12: Striation motion. Ionization takes place in the region between the maximum of the electric field and of the plasma density leading the generation of electron-ion pairs by ionization ahead of the maximum plasma density in the electric field direction i.e. in the cathode direction (to the left of the figure). The maximum plasma density therefore moves in this direction.

The phase velocity of the striations can be easily measured in the simulations for example by looking at two-dimensional contours of plasma density (or other quantities) in the position-time, $(x, t)$ plane. This is illustrated in Figure 13, which displays a contour plot of the space and time variations of the plasma density for two values of the applied electric field., $E_0 = 6$ V/cm and $E_0 = 17$ V/cm in neon at $p = 1$ torr and for a simulated (periodic) length $L = 10$ cm.

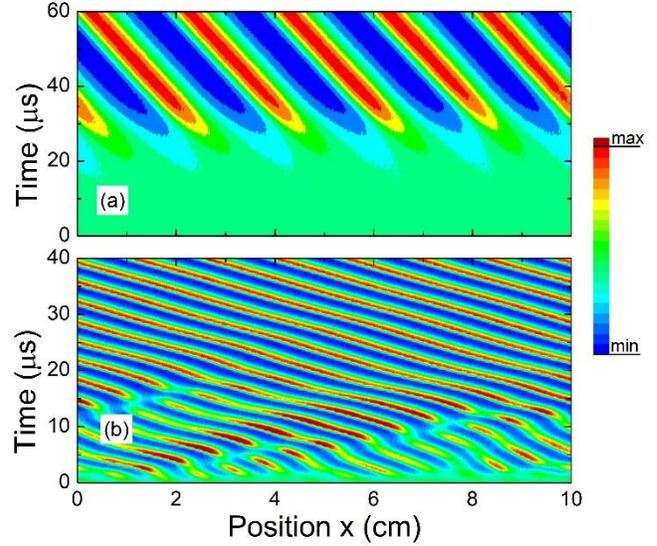

Figure 13: Two-dimensional $(x, t)$ contours of constant plasma density in neon, 1 torr for a periodic length $L = 10$ cm of the simulation domain, for an applied field a) $E_0 = 6$ V/cm , and b), $E_0 = 17$ V/cm (usual pressure scaling laws can be used). [min, max] are the minimum and maximum plasma densities in the non-linear regimes and are respectively [$0.44 \times 10^{15}$ m$^{-3}$, $1.78 \times 10^{15}$ m$^{-3}$] for the $E_0 = 6$ V/cm case, and [$0.2 \times 10^{15}$ m$^{-3}$, $2.2 \times 10^{15}$ m$^{-3}$] for the $E_0 = 17$ V/cm case. The initial conditions, at time $t = 0$ correspond to a uniform, neutral plasma of density $10^{15}$ m$^{-3}$.

From these plots, we can infer that the velocity $v_s$ of the striations is about 0.7 km/s for $E_0 = 6$ V/cm and 4 km/s for $E_0 = 17$ V/cm. The wavelengths are respectively 2 cm and 1.4 cm for the two cases, so the wave frequencies $F_s = v_s/\lambda$ are respectively on the order of 35 kHz and 300 kHz. More generally, the striation velocity increases with increasing field, from about 200 m/s for the lowest values of the field considered in the simulations for neon, to more than 5 km/s for the largest values of the electric field (i.e. between $E_0 = 4.5$ V/cm and $E_0 = 30$ V/cm), and the striation frequency varies from 7 kHz to 500 kHz. This order of magnitude is consistent with the experimental values[2] although detailed comparisons with experiments are difficult because stepwise ionization is not always negligible in reported experiments.

It is interesting to note that, in Figure 13b, for $E_0 = 17$ V/cm, the instability forms with 10 maxima (i.e. a wavelength of 1 cm since the period is $L = 10$ cm) in the linear phase and evolves to a steady state regime with 7 maxima (wavelength 1.4 cm) in the non-linear phase. For $L = 6$ cm and the same applied field, the wavelength was about 1 cm (see Figure 2b and Figure 14). This indicates that the wavelength in the non-linear, steady state regime is not necessarily identical to the wavelength that has the maximum growth rate in the linear phase. In most simulations with a given length $L$ of the periodic simulation domain, we observed, in the non-linear phase very regular, monochromatic waves as can be seen in Figure 13.

## IV. LINEAR STABILITY ANALYSIS AND COMPARISON WITH THE PIC-MCC RESULTS

In this section we perform a linear stability analysis based on a three moment description of electron transport in the quasineutral plasma, using the same approach as Désangles et al.[24] who analyzed the development of striations observed in an annular, inductively coupled radiofrequency argon plasma.

It is clear that the fluid model cannot describe the spatial resonances described in the literature and observed in the PIC-MCC simulations above. These resonances are associated with the discrete nature of the energy losses of electrons due to excitation and ionization and can only be described by kinetic equations. However we think that the linear stability analysis of the fluid model can predict the conditions of instability formation and can provide a clear understanding of the physics responsible for triggering the instability. We also show in this section that despite the limitations of the fluid model, the growth rate, and the wavelength and frequency of the instability at the maximum growth rate predicted by the linearized fluid model, and their variations with the applied field follow relatively closely the trends observed in the kinetic PIC-MCC simulations.

This linear stability analysis is presented in section IV.A. The domain of existence, growth rate and wavelength of the instability predicted by this stability analysis are compared section IV.B with those deduced from the PIC-MCC model.

### E. Fluid model equations

We consider one-dimensional (longitudinal) fluid equations for a quasineutral plasma where the transport coefficients are supposed to depend on the local electron temperature as obtained from solutions of the 0D Boltzmann equation. The equations below are a fluid approximation of the PIC-MCC model.

The electron continuity and energy equation are written as in Ref.[24]:

$$\partial_t n_e + \partial_x \Gamma_e = n_e(\nu_i - \nu_w) \tag{3}$$

$$\partial_t (3/2 n_e T_e) + \partial_x H_e = P - n_e \mathcal{L}_\varepsilon \tag{4}$$

$n_e$ is the electron density, $\Gamma_e$ the electron flux and $H_e$ the electron energy flux. $T_e$ is the electron temperature (in eV) defined by $T_e = 2/3\bar{\varepsilon}_e$ ($\varepsilon_e$ is the electron mean energy). $\nu_i$ is the ionization frequency and $\mathcal{L}_\varepsilon$ (in units of eV/s) is the energy loss per unit time due to electron-neutral collisions. $\nu_w$ is the frequency of charged particle losses to the wall. $P$ is the power density (in eV/m³/s) absorbed by the electrons and is equal to:

$$P = -\Gamma_e E \tag{5}$$

The ionization and the energy loss frequencies are assumed to depend on the electron temperature with the same functional dependance as under uniform field conditions. $\nu_i(T_e)$ and $\mathcal{L}_\varepsilon(T_e)$ can be obtained and tabulated by solving the steady state homogeneous Boltzmann equation for different values of the reduced electric as in Figure 7. The BOLSIG+ solver[42] was used in this paper to obtain $\nu_i(T_e)$ and $\mathcal{L}_\varepsilon(T_e)$. $\mathcal{L}(T_e)$ is sometimes written as[45]:

$$\mathcal{L}_\varepsilon(T_e) = \nu_i(T_e) \, \mathcal{E}_c(T_e) \tag{6}$$

where $\mathcal{E}_c$ is the energy loss per electron-ion pair created.

The electron flux and electron energy flux are written as in Ref.[24], in a drift-diffusion form:

$$\Gamma_e = -n_e \mu_e E - \partial_x (n_e D_e) \tag{7}$$

$$H_e = -n_e \beta_e E - \partial_x (n_e G_e) \tag{8}$$

$\mu_e$ and $D_e$ are the electron mobility and diffusion coefficients. $\beta_e$ and $G_e$ are the thermoelectricity and heat diffusion transport coefficients[46] and are related to the electron energy mobility and electron energy diffusion coefficients, $\mu_\varepsilon$ and $D_\varepsilon$, by:

$$\beta_e = 3/2 T_e \mu_e \quad \text{and} \quad G_e = 3/2 T_e D_e \tag{9}$$

The electron temperature dependent transport coefficients $\mu_e(T_e)$, $D_e(T_e)$, $\mu_\varepsilon(T_e)$ and $D_\varepsilon(T_e)$ are also obtained from a tabulation of BOLSIG+ solutions of the 0D Boltzmann equation[42].

The charged particle current density in the plasma column, $J_T$, must be constant in space so the total flux $\Gamma_T = J_T/e = \Gamma_i - \Gamma_e$ is a constant. Neglecting ion diffusion and using the quasineutrality condition $n_i = n_e$, the electric field and the ion flux $\Gamma_i$ are related by:

$$E = \Gamma_i/(n_e \mu_i) = (\Gamma_e + \Gamma_T)/(n_e \mu_i) \tag{10}$$

Using equation (10), the electric field can be eliminated from the electron flux. This gives:

$$\Gamma_e = -\mu_i/(\mu_e + \mu_i) \, \partial_x (n_e D_e) - \mu_e/(\mu_e + \mu_i) \, \Gamma_T \tag{11}$$

If we assume $\mu_i \ll \mu_e$ and a weak dependence of the mobility coefficients on the electron temperature, the first term on the right hand side of eq. (11) is the usual ambipolar diffusion term $-\partial_x (n_e D_a)$ and the ambipolar diffusion coefficient $D_a = \mu_i D_e/\mu_e$ is a function of electron temperature. The second term is a constant independent on position. We can therefore express $\partial_x(n_e D_a)$ as:

$$\partial_x (n_e D_a) = D_a \partial_x n_e + n_e (\mu_i/\mu_e) \partial_{T_e} D_e \partial_x T_e$$

Using similar approximations for the electron flux and the electron energy flux, we finally obtain, using the notations of Désangles et al.[24]:

$$\Gamma_e = -D_a \partial_x n_e - \eta_e \partial_x T_e - \mu_e/(\mu_e + \mu_i) \, \Gamma_T \tag{12}$$

$$H_e = -\chi_e \partial_x n_e - \kappa_e \partial_x T_e - \beta_e/(\mu_e + \mu_i) \, \Gamma_T \tag{13}$$

The coefficient of the density and temperature gradients in eqs. (12) and (13) are:

$$\eta_e = n_e (\mu_i/\mu_e) \partial_{T_e} D_e \tag{14}$$

$$\kappa_e = n_e (\partial_{T_e} G_e - \beta_e/\mu_e \partial_{T_e} D_e) \tag{15}$$

$$\chi_e = G_e - (\beta_e/\mu_e) D_e \tag{16}$$
$$= 3/2 T_e (D_\varepsilon - (\mu_\varepsilon/\mu_e) D_e)$$

Equations (12) and (13) are identical to those of Ref.[24] except for the last term of the right hand side, which is related to the constant dc current in the plasma column (these terms are not present in Ref.[24] because a radiofrequency field was considered in this paper so the time averaged current was equal to zero).

Combining eqs. (10) and (12) gives the electric field:

$$E = -\partial_x (n_e D_e/\mu_e)/n_e + \Gamma_T/[n_e(\mu_e + \mu_i)] \tag{17}$$

The first term on the right-hand side of eq. (17) is the ambipolar component of the electric field and the second term is the resistive component.

The solutions of the continuity and energy equations (3) and (4) corresponding to a steady state, homogeneous plasma column of density $n_0$, electron temperature $T_0$ and electric field $E_0$ are obtained by writing that the right hand side of these equations is zero (the subscript 0 indicates the homogenous, steady state regime), i.e.:

$\nu_{i0} \equiv \nu_i(T_0) = \nu_w$

$P_0 = -\Gamma_{e0} E_0 = \Gamma_{e0}^2/(n_0 \mu_{e0}) = n_0 \mathcal{L}_{\varepsilon0}$

Note that for a cylindrical plasma column of radius $R$ where ambipolar diffusion to the walls is the only charged particle loss mechanism, we can write as above $\nu_{i0}(T_0) = \nu_w = D_{a0}(T_0)/(2.4R)^2$. This equation defines the electron temperature $T_0$ (and corresponding electric field $E_0$) in the homogeneous plasma column.

### F. Linear stability analysis

We consider small perturbations of the electron density and electron temperature, $\delta n_e$ and $\delta T_e$, around the uniform solution $(n_0, T_0)$. We look for solutions of the fluid equations (3) and (4) of the form:

$n_e = n_0 + \delta n_e e^{-i\Omega t + ikx}, \ T_e = T_0 + \delta T_e e^{-i\Omega t + ikx}$

with $\Omega = \omega + i\gamma$. $\omega$ is the angular frequency, $\gamma$ is the growth rate and $k$ the wave number of the perturbation. Instabilities form if the growth rate $\gamma$ is positive.

The linearization of the collision terms of the continuity and energy equation gives:

$\nu_i = \nu_{i0} + \partial_{T_e}\nu_i\big|_{T_0} \delta T_e, \ \mathcal{L}_\varepsilon = \mathcal{L}_{\varepsilon0} + \partial_{T_e}\mathcal{L}_\varepsilon\big|_{T_0}\delta T_e$

Using eq. (17), the linearization of the electric field gives:

$\delta E = -(E_0 + ik\,D_{e0}/\mu_{e0})\,\delta n_e/n_0$

(the $\delta T_e$ term is neglected) and the electron heating term $P = -\Gamma_e E$ of the energy equation can be written as:

$P = n_0 \mathcal{L}_{\varepsilon0} - \mathcal{L}_{\varepsilon0}\delta n_e - ik E_0 D_{e0}\delta n_e$

Finally, linearizing eqs. (3) and (4) with respect to $\delta n_e$ and $\delta T_e$, we obtain, after a few approximations:

$$\begin{bmatrix} -i\omega + \gamma + D_{a0}k^2 & -n_0\nu'_{i0} \\ \chi_{e0}k^2 + 2\mathcal{L}_{\varepsilon0} + ikE_0 D_{e0} & \kappa_{e0}k^2 + n_0\mathcal{L}'_{\varepsilon0} \end{bmatrix} \begin{bmatrix} \frac{\delta n_e}{n_0} \\ \frac{\delta T_e}{T_e} \end{bmatrix} = \begin{bmatrix} 0 \\ 0 \end{bmatrix} \quad (18)$$

where we have used the notations:

$\nu'_{i0} = \partial_{T_e}\nu_i\big|_{T_0}$ and $\mathcal{L}'_{\varepsilon0} = \partial_{T_e}\mathcal{L}_\varepsilon\big|_{T_0}$

Some of the approximations leading to eq. (18) are: the term $k^2\eta_e$ is neglected compared to $n_0\nu'_{i0}$ in the top right element of the matrix; the $\partial_t$ term of the energy equation is neglected. The validity of these approximations was checked *a posteriori*.

Writing that the determinant of the matrix of eq. (18) must be zero, we get approximate expressions of the angular frequency and growth rate, $\omega$ and $\gamma$:

$$\omega = \frac{k\nu'_{i0}E_0 D_{e0}}{k^2 \kappa_{e0}/n_0 + \mathcal{L}'_{\varepsilon0}} \quad (19)$$

$$\gamma = -D_{a0}k^2 - \nu'_{i0}\frac{2\mathcal{L}_{\varepsilon0} + \chi_{e0}k^2}{k^2 \kappa_{e0}/n_0 + \mathcal{L}'_{\varepsilon0}} \quad (20)$$

These expressions are independent of $n_0$ since $\kappa_{e0}$ is proportional to $n_0$. In the following we use $K_{e0} = \kappa_{e0}/n_0$.

The first term of the equation for the growth rate $\gamma$ is negative and is a stabilizing term due to ambipolar diffusion. The second term must be sufficiently positive for instabilities to grow which is possible only if $\chi_{e0}$ is sufficiently negative

The maximum value $\gamma_{max}$ of the growth rate as a function of the wave number and the wave number $k_{max}$ at the maximum growth rate can be deduced from the above expression of $\gamma$:

$$\gamma_{max} = D_{a0}\big[\beta - \alpha - 2(\delta - \alpha\beta)^{1/2}\big] \quad (21)$$

$$k_{max} = \big[-\beta + (\delta - \alpha\beta)^{1/2}\big]^{1/2} \quad (22)$$

With $\alpha = \frac{\chi_{e0}\nu'_{i0}}{D_{a0}K_{e0}}, \beta = \frac{\mathcal{L}'_{\varepsilon0}}{K_{e0}}, \delta = \frac{2\mathcal{L}_{\varepsilon0}\nu'_{i0}}{D_{a0}K_{e0}}$

The maximum growth rate is positive, i.e. instabilities can develop if and only if $\alpha < -\beta - 2\sqrt{\delta}$, which gives:

$$\chi_{e0} < -\frac{D_{a0}\mathcal{L}'_{\varepsilon0} + \sqrt{8D_{a0}K_{e0}\mathcal{L}_{\varepsilon0}\nu'_{i0}}}{\nu'_{i0}} \quad (23)$$

This condition is identical to the condition derived by Désangles et al.[24] (eq. (10) of this reference) except for the numerical factor 8 instead of 6 under the square root. This difference is due to a slightly different assumption on the heating term of the energy equation (inductive radiofrequency heating was considered in Ref.[24]).

The inequality (23) indicates that the instability develops if $\chi_{e0}$ is sufficiently negative, i.e., from the definition of $\chi_e$, eq. (16), if $D_\varepsilon - (\mu_\varepsilon/\mu_e)D_e$ is sufficiently negative. Since $\mu_\varepsilon/\mu_e$ is often not far from 5/3 (the value for a Maxwellian distribution)[42], this means that the space diffusion coefficient $D_e$ must be sufficiently larger than 3/5 of the energy diffusion coefficient, $3/5D_\varepsilon$. This is possible (see Fig. 9a of Ref.[42]), e.g. for the non-Maxwellian electron energy distribution functions of inert gases under a uniform field.

The maximum growth rate $\gamma_{max}$ and corresponding wave number $k_{max}$ are functions of the electron temperature $T_0$. The coefficients $\alpha, \beta, \delta$ can be obtained and tabulated as a function of the electron temperature $T_0$ from solutions of the 0D Boltzmann equation for different values of $E_0$. The phase velocity of the wave at the maximum growth rate can be deduced from the angular frequency and wave number at the maximum growth rate: $v_{ph,max} = \omega_{max}/k_{max}$.

### G. Comparison with the PIC-MCC model

The growth rate, wave number and angular frequency or phase velocity of the instability can be deduced from the results of the PIC-MCC simulation, and their variations with the reduced electric field $E_0/p$ can be compared with those given by the linear analysis of the fluid model, eqs. (21) and (22). These comparisons are shown in Figure 14 and Figure 15 for the wavelength and growth rate respectively. Simulations with different values of the periodic length are shown for the case of neon ($pL = 4, 6, 10$, and 15 torr. cm). As said above, since stepwise ionization is neglected, the usual pressure scaling laws can be used and the results are presented in reduced units (the PIC-MCC simulations have been performed at 1 torr).

As mentioned at the beginning of this section the fluid model cannot predict the spatial resonances of the $p$, $r$, and $s$ waves, and the maximum growth rate and wavelength at the maximum growth rate deduced from the linearized fluid model are not identical to the growth rate and wavelength observed in the non-linear regime of the PIC-MCC simulations. Nevertheless, it is interesting to note on Figure 14 and Figure 15 that the values of these parameters and their variations with $E_0/p$ deduced from the fluid and PIC-MCC models are relatively close.

The linear analysis predicts that there is no instability for $E_0/p < 3.2$ V/cm/torr in neon and $E_0/p < 8$ V/cm/torr in argon. No instabilities were found with the PIC-MCC model for $E_0/p$ values below 4.3 and 11 V/cm/torr and for simulation times up to $10^{-3}$ s.

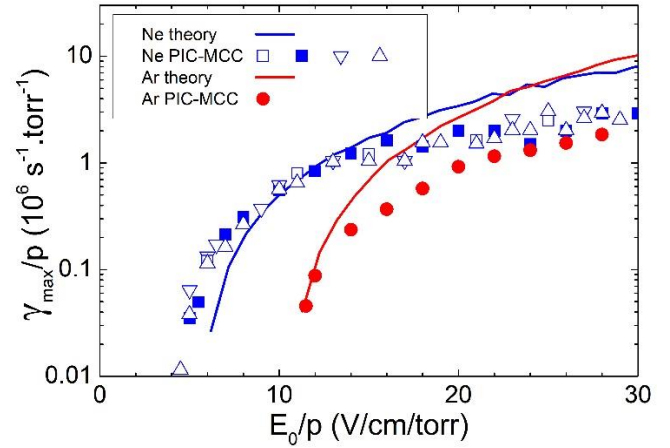

Figure 15: Growth rate of the instability in neon and argon as a function of $E_0/p$ from the linear stability analysis (eq. (22) full lines, and from PIC-MCC simulations (symbols). PIC-MCC results obtained in the non-linear steady state regime with different values of the periodic length $L$ of the simulated domain are shown (full symbols, $pL = 6$ torr. cm ; open up-triangles neon with $pL = 15$ torr. cm ; open down-triangles, neon with $pL = 10$ torr. cm ; open rectangles, neon with $pL = 4$ torr. cm).

For larger electric fields, a second branch of the $(p\lambda, E_0/p)$ clearly appears in the PIC-MCC simulations both in neon and argon.

According to the values of the potential drop between striations calculated from the PIC-MCC simulations and shown in Figure 16 the first branch of the $(p\lambda, E_0/p)$ plot of Figure 15 corresponds to $r$ ionization waves (potential drop of 12-14 V in neon in the experiments of Golubovsky et al.[14], around 13 V in the PIC-MCC simulations) and may include $p$ waves for the lowest field values in neon (potential drop on the order of 10 V in neon), while the second branch correspond to $s$ waves (potential drop of 17-22 V in neon in the experiments of Golubovsky et al.[14], around 24 V in the PIC-MCC simulations).

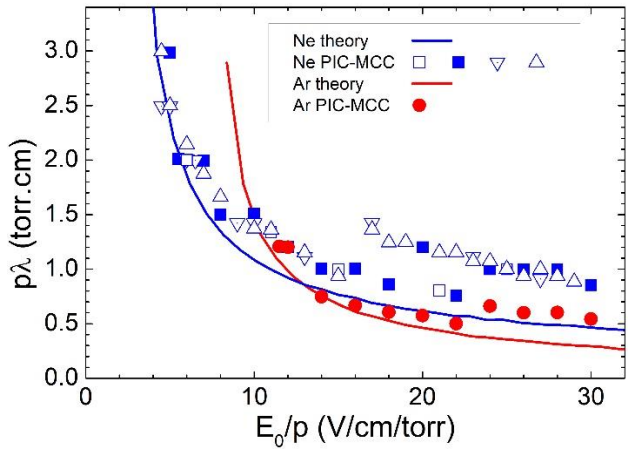

Figure 14: Wavelength of the instability in neon and argon as a function of the reduced electric field $E_0/p$ from the linear stability analysis ($\lambda = 2\pi/k_{max}$ with $k_{max}$ from eq. (22), full lines) and from PIC-MCC simulations (symbols). PIC-MCC results obtained with different values of the periodic length $L$ of the simulated domain are shown (full symbols, $pL = 6$ torr. cm, open up-triangles neon with $pL = 15$ torr. cm, open down-triangles, neon with $pL = 10$ torr. cm, open rectangles, neon with $pL = 4$ torr. cm). The PIC-MCC wavelength shown in this plot was measured in the non-linear, steady state regime of the ionization wave.

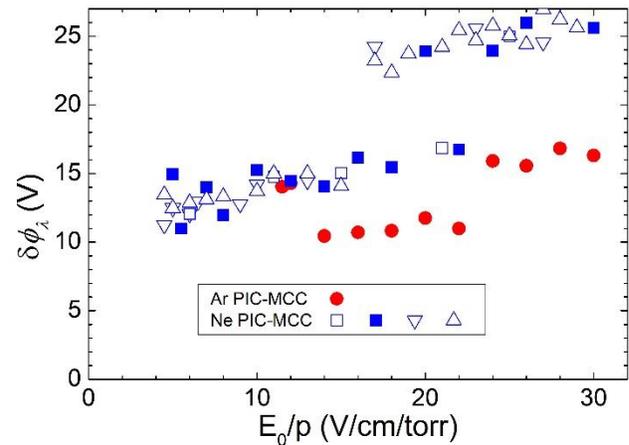

Figure 16 : Potential drop along a striation length as a function of $E_0/p$ from the PIC-MCC simulations (full symbols, $pL = 6$ torr. cm ; open up-triangles neon with $pL = 15$ torr. cm ; open down-triangles, neon with $pL = 10$ torr. cm ; open rectangles, neon with $pL = 4$ torr. cm).

Finally, we compare in Figure 17 the phase velocity of the wave predicted by the linear stability analysis, with the phase velocity deduced from the PIC-MCC simulations. The phase velocity increases from a few 100 m/s to a few km/s in the considered range of average electric field. Again, we note the the relative good agreement between predictions of the fluid model and PIC-MCC results in spite of the limitations of the fluid model.

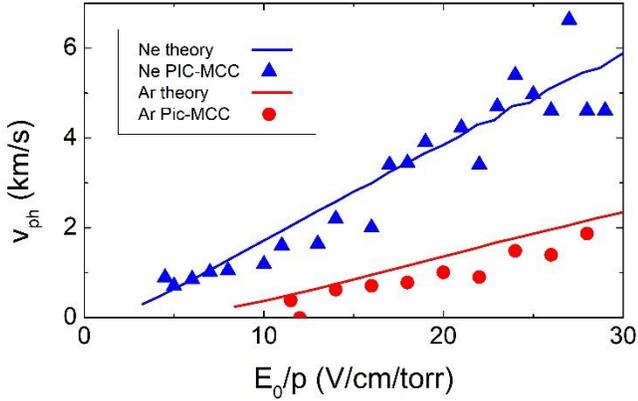

Figure 17 : Phase velocity of the ionization wave at the maximum growth rate from the fluid model ($v_{ph} = \omega / k$ with $\omega$ from eq. (19), calculated for $k = k_{max}$ , eq. (22) ; full lines), compared with the phase velocity deduced from PIC-MCC simulations in the non-linear regime (triangle symbols : neon with $pL = 15$ torr.cm , circle symbols : argon with $pL = 6$ torr. cm ).

## V.   REMARKS ON THE MODEL: PERIODIC VS BOUNDED PLASMA MODEL

In the model used in this paper and described in section II, the average electric field in the column was imposed and periodic boundary conditions were used. Wall losses were described by removing an electron and an ion when an electron-ion pair was generated by ionization. This method allowed to keep the average plasma density (and the number of particles in the simulation) constant in time, and was very convenient for measuring the growth rate of the instability and comparing with the predictions of the linearized fluid model.

Other methods are of course possible. For example, it is possible to impose a charged particle loss frequency instead of imposing the average electric field. In that case, the electric field has to adjust in order to satisfy balance between ionization and wall losses. This can be done by iterations or by adding an external circuit with a resistor to adjust the applied electric field $E_0$ in the case of periodic boundary conditions, or to control the applied voltage in the case of a bounded discharge. In the latter case, the cathode and anode regions are described and it may be interesting to study the effects of these boundaries on the development and properties of the ionization waves. The detailed description of a cathode region controlled by secondary electron emission and ionization may be cumbersome in a PIC-MCC simulation because of the large plasma density in the negative glow. A way to simplify the treatment of the cathode region is to impose an electron current emitted by the cathode (instead of considering secondary electron emission by ion impact). This regime corresponds to a glow discharge with a hot cathode.

We performed PIC-MCC simulations based on this model of a bounded plasma. The results concerning the wavelength of the ionization waves and the potential drop along a striation length are very similar to those obtained with periodic boundary conditions. An interesting feature is the oscillations of the anode sheath electric field in phase with the development of the ionization wave, i.e. the successive formation of an electron sheath and of an ion sheath at the wave frequency, in front of the anode. Figure 18-19 shows some results for a neon discharge at $p = 1$ torr, with a cathode-anode gap of $L = 15$ cm, an applied voltage of $U = 270$ V, an electron emission current density of $J_{e0} = 10$ A/m² at the cathode, a resistor $R = 10$ Ω. m² in the external circuit, and a fixed charged particle loss frequency of $10^6$ s⁻¹.

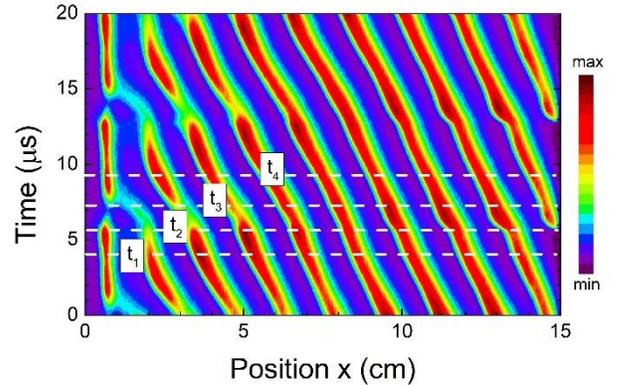

Figure 18: Contour plot of the electron density for a bounded plasma column in neon as a function of space and time over a time interval of 20 µs in the steady state regime. The cathode is on the left side, $p = 1$ torr, $L = 15$ cm. A voltage of 270 V is applied between cathode and anode through a resistor $R = 10$ Ω. m² and an electron emission current density of $J_{e0} = 10$ A/m² is emitted at the cathode. Charged particles are lost to the walls at a frequency of $10^6$ s⁻¹. The calculated discharge current is 6 A/m², corresponding to a cathode-anode voltage of 210 V. The [min, max] values of the electron density are [$10^{13}$, $1.4 \times 10^{15}$] m⁻³. PIC-MCC simulations with 1000 grid points and 5000 particles per cell.

The calculated total current density in these conditions is $J_T = 6$ A/m², i.e. lower than the cathode emitted electron current density, which indicates that electron emission is space charged limited (as in a hot cathode discharge). The cathode-anode voltage is therefore $V_D = U - R\,J_T = 210$ V. Figure 18 displays a contour map of the calculated electron density as a function of position and time over about three cycles of the ionization wave. We can see the formation of cathode-directed ionization waves at a frequency close to 150 kHz. There are $m = 9$ maxima of the electron density so the average wavelength is $\lambda = L/m \approx 1.7$ cm and the potential drop along one striation is $\delta_{\phi} = V_D/m \approx 23.3$ V , which corresponds to an $s$ wave. The average electric field $E_0 = V_D/L$ is 14 V/cm.

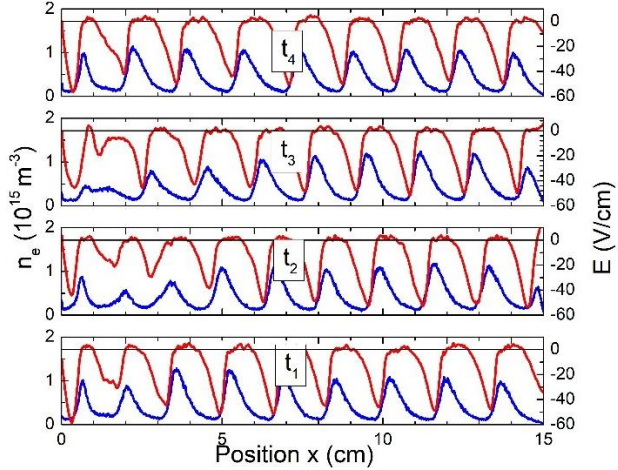

Figure 19: Axial profiles of the electron density and electric field in the conditions of Figure 18 at four different times $t_1$, $t_2$, $t_3$, $t_4$ indicated on Figure 18.

Figure 19 shows the profiles of the electron density and electric field at four different times, $t_1$, $t_2$, $t_3$, $t_4$ indicated on Figure 18. We see on Figure 18 and Figure 19 the generation of new plasma density maxima at the anode, and how the wave connects to the non-moving density maximum in the cathode region (the interaction of the wave with the cathode region depends on the structure of this region and would be different in the conditions of a standard, cold cathode glow discharge).

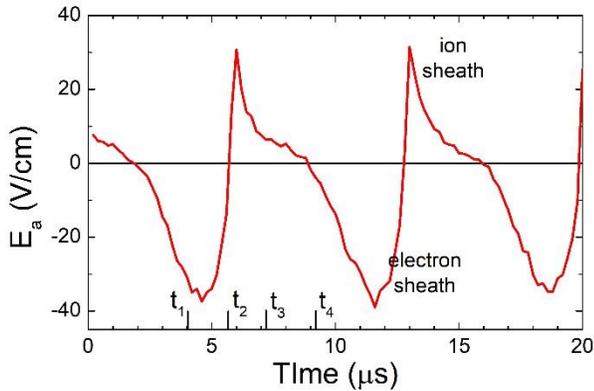

Figure 20: Time evolution of the anode electric field in the conditions of Figure 18 and Figure 19.

As said above , the electric field changes sign periodically on the anode side, i.e. the anode sheath is alternatively an electron sheath (times $t_1$ and $t_4$) and an ion sheath (times $t_2$ and $t_3$). The oscillations of the anode electric field $E_a$ displayed in Figure 20 exhibit sharp transitions between the electron sheath and ion sheath regimes. Such oscillations of the anode electric field have been observed in experimental work[47] and recently predicted in the self-consistent hybrid model of kinetic striations of Kolobov et al.[23].

## VI. CONCLUSION

A self-consistent 1D PIC-MCC model was used to study the development of instabilities in positive column plasmas in inert gases under conditions where stepwise ionization and Coulomb collisions are negligible and where charged particle losses are due to ambipolar diffusion to the walls. The model is able to reproduce satisfactorily (and more than qualitatively) most of the experimentally observed properties of the instabilities known as ionization waves or moving striations of the $p$, $r$, and $s$ types in positive column plasmas in inert gases. $r$ and $s$ types were clearly apparent in the simulations while the prediction of $p$ waves was not as obvious. The wavelengths of the ionization waves, potential drop along a striation length and phase velocity of the waves predicted by the PIC-MCC simulations are consistent with the experiments and with theories invoking the existence of kinetic resonances controlling the potential drop and wavelength of the striations.

The plasma density gradients in the striations generate a large ambipolar electric field and the plasma density oscillations in space are coupled with strong oscillations of the electric field with a phase-shift imposed by current continuity. Electrons are trapped in the low field regions around the plasma density maxima and the electron energy probability function (EEPF) presents a large population of low energy electrons in these regions. These electrons are accelerated by the large electric field between plasma density maxima, leading to "bumps" ("electron bunching" in the literature[4, 12]) in the EEPF at energies corresponding to the potential drop between striations.

Although the properties of electron transport in the striated (non-linear) regime are clearly kinetic in nature and cannot be described by a fluid model, we find that the mechanisms triggering the instabilities can be very well described by a three-moments fluid model taking into account the non-Maxwellian nature of the EEPF in the uniform field prior to the formation of the instability and in the linear phase of the instability development. From a linear stability analysis of this fluid model, it appears that an essential feature controlling the existence of instabilities leading to ionization waves of the $p$, $r$, $s$ types in the considered conditions is the particular shape of the EEPF in the unperturbed electric field that leads to electron diffusion in space sufficiently larger than electron diffusion in energy (as was concluded in the recent paper of Désangles et al.[24] on instabilities in an inductive rf discharge).

The evolution toward the non-linear, saturated state regime in the simulations is controlled by the kinetic resonances observed in the experiments and predicted by previous theoretical and numerical works. The linearized fluid model cannot describe the spatial resonances but its predictions of the variations of the maximum growth rate, wavelength and wave frequency at the maximum growth rate with applied field are, interestingly, not very far from the results of the kinetic simulations.

The results presented in this paper assume negligible stepwise ionization. Stepwise ionization may actually contribute to the total ionization in a non-negligible way under some of the conditions considered in this paper. It would be useful to check how stepwise ionization would modify the results of the PIC-MCC simulations and the predictions of the linear stability analysis of the fluid model.

## ACKLNOWLEDGMENTS

The author would like to thank G.J.M. Hagelaar and L.C. Pitchford for useful discussions and comments.

## DATA AVAILABILITY

The data that support the findings of this study are available from the corresponding author upon reasonable request.